\documentclass[12pt]{iopart}

\usepackage{iopams}  
\usepackage{graphicx}
\usepackage{bm}
\expandafter\let\csname equation*\endcsname\relax
\expandafter\let\csname endequation*\endcsname\relax
\usepackage{amsmath}
\usepackage{epsfig}
\usepackage{rotating}
\usepackage{cite}

\def\pr{\prime}
\def\be{\begin{equation}}
\def\lan{\left\langle}
\def\ran{\right\rangle}
\def\ee{\end{equation}}
\def\barr{\begin{array}}
\def\earr{\end{array}}

\def\l{\left}
\def\r{\right}
\def\dis{\displaystyle}
\def\ed{\end{document}}
\def\f{\frac}

\def\cam{{\cal M}}

\def\ek{{\hat{E}_\kappa}}

\begin{document}

\title[Conditional $q$-normal form of strength functions in many-fermion systems]{Wavefunction structure in quantum many-fermion systems with $k$-body interactions: conditional $q$-normal form of strength functions}

\author{V. K. B. Kota}
\ead{vkbkota@prl.res.in}
\address{Physical Research Laboratory, Ahmedabad 380 009, India}
\author{Manan Vyas}
\ead{corresponding author, manan@icf.unam.mx}
\address{Instituto de Ciencias F{\'i}sicas, Universidad Nacional 
Aut{\'o}noma de M\'{e}xico, 62210 Cuernavaca, M\'{e}xico}

\begin{abstract}

For finite quantum many-particle systems modeled with say $m$ fermions in $N$ single particle states and interacting with $k$-body interactions ($k \leq m$), the wavefunction structure is studied using random matrix theory. Hamiltonian for the system is chosen to be $H=H_0(t) + \lambda V(k)$ with the unperturbed $H_0(t)$ Hamiltonian being a $t$-body operator and $V(k)$ a $k$-body operator with interaction strength $\lambda$. Representing $H_0(t)$ and $V(k)$ by independent Gaussian orthogonal ensembles (GOE) of random matrices in $t$ and $k$ fermion spaces respectively, first four moments, in $m$-fermion spaces, of the strength functions $F_\kappa(E)$ are derived; strength functions contain all the information about wavefunction structure. With $E$ denoting the $H$ energies or eigenvalues and $\kappa$ denoting unperturbed basis states with energy $E_\kappa$, the $F_\kappa(E)$ give the spreading of the $\kappa$ states over the eigenstates $E$. It is shown that the first four moments of $F_\kappa(E)$ are essentially same as that of the conditional $q$-normal distribution given in: P.J. Szabowski, Electronic Journal of Probability {\bf 15}, 1296 (2010). This naturally gives asymmetry in $F_\kappa(E)$  with respect to $E$ as $E_\kappa$ increases and also the peak value changes with $E_\kappa$. Thus, the wavefunction structure in quantum many-fermion systems with $k$-body interactions follows in general the conditional $q$-normal distribution. 

\end{abstract}


\maketitle

\section{Introduction}

Wavefunction structure in finite quantum many-body systems follows from the form of the strength functions and its parameters. Given the eigenstates expanded in terms of a set of physically motivated basis states, strength functions correspond to the spread of a basis state over the eigenstates. More importantly, they determine the chaos measures in generic many-body systems - number of principle components (NPC) and information entropy ($S^{info}$) in wavefunctions \cite{KS, Ko-book, Bo-17, Ga-18, Vil-20, To-17, Ma-14, Ri-08, Ri-16}. NPC gives the number of basis states that make up the eigenstate and $S^{info}$ is a measure of entropy in the eigenstate. In addition, strength functions also determine fidelity decay, out-of-time order correlator (OTOC) and many other aspects of wavefunctions, which are essential to understand non-equilibrium dynamics of isolated finite complex quantum systems  \cite{HMKC, Zelea, Ta-16, PRR-20}. OTOC is also useful in information scrambling \cite{Swin-18, PRR-20, Arul-19, Sch-19, Mal-16}.

Strength functions, also known as Local Density of States (LDOS), are an important quantity in studying dynamics of a finite many-particle system \cite{Ko-book, Zelea, Bo-17, Ga-18, Vil-20, To-17, Ma-14}. Considering the quench dynamics described by a Hamiltonian $H = H_0 + \lambda V(k)$, we prepare the system in unperturbed eigenstates of $H_0$ and study how these states spread in the unperturbed many-body basis due to $V(k)$. The strength functions describe the average energy distribution of the initial states by projecting them on the energy eigenbasis (note that in practice an averaging over the initial states, chosen in an energy bin, is also carried out). It essentially gives the intensity with which an eigenstate is contained in unperturbed basis of the total Hamiltonian. As a function of increasing interactions (with a one-body unperturbed part (i.e. $t = 1$ and $k = 2$), the strength functions make a crossover from delta function (non-interacting regime) to Breit-Wigner distribution (localized regime) to Gaussian distribution (chaotic/thermodynamic regime) \cite{Ko-book}. The widths of the strength function determine the decay rate of NPC \cite{Lea-PRE-R} and OTOC \cite{Lea-19} for quenched bosonic systems in the regime of strong chaos. Thus, the structure of strength functions affects many-body system dynamics and is an essential ingredient in understanding wavefunction structure of quantum many-body systems, in close connection with the problem of thermalization in generic many-body systems. Let us mention that earlier studies on strength functions from the point of view of quantum chaos and random matrix theory are due to Flambaum, Izrailev, Shepelyansky, Zelevinsky and many others \cite{fke-e1, fke-e2, fke-e3, fke-e4, fke-e5, fke-e6, fke-e7}.

For a finite fermion or boson system with the particles in a mean-field and interacting with two-body interactions, it is well established that in the strong coupling limit (or in the thermodynamic region, i.e. the region where different definitions of entropy, temperature etc. give the same results or equivalently the region where usual thermodynamic principles apply [2, 9, 11]) strength functions follow Gaussian form \cite{Ko-book,Zelea}. This result extends to the situation with $k$-body interactions for $k$ much less than number of particles. In this paper we will present results, obtained using random matrix theory, for strength functions valid for any $k$; note that $k$ is less than or equal to the number of particles. For many-particle systems with $k$-body interactions, the appropriate random matrix ensembles are $k$-body embedded ensembles \cite{Ko-book}. One very important property of these ensembles is that the form of the eigenvalue density for a $m$ particle system with $k$-body interactions is Gaussian for $k << m$ and semi-circle for $k=m$ \cite{MF,Br-81}.  Sachdev-Ye-Kitaev models are also examples of embedded ensembles with complex fermions replaced by Majorana fermions and have been receiving increasing attention in high-energy physics \cite{Ki-1, Ki-2, Sa-93, Gu-20, Ver-a, Ver-b, Verba-1, Ver-d}.  

Recently, a new direction in exploring embedded ensembles has opened up with the recognition that the eigenvalue density (ignoring fluctuations) is given by the so-called $q$-normal distribution $f_{qN}(x|q)$ generating correctly Gaussian form for the parameter $q$ taking the value $q=1$ and semi-circle for $q=0$ \cite{MK}. The $q$-normal distribution is related to $q$-Hermite polynomials that reduce to normal Hermite polynomials for $q=1$ and Chebyshev polynomials for $q=0$ \cite{Ismail}. The embedded ensembles and $q$-normal correspondence follows from the novel results obtained by Verbaarschot for quantum black holes with Majorana fermions \cite{Verba-1,Verba-2}. In another important recent development \cite{MK-new}, it is shown that the bivariate $q$-normal distribution $f_{biv-qN}$ defined in \cite{Sza-1} gives the form for the bivariate transition strength densities generated by a $k$-body Hamiltonian represented by embedded ensembles with the transition operator represented by an independent embedded ensemble. With these investigations, clearly $q$-normal and bivariate $q$-normal are expected to be useful in describing strength functions generated by $k$-body embedded ensembles. 

In \cite{MK}, it is shown using numerical calculations with both fermion and boson systems and $H_0$ representing a mean-field one-body part, that the strength functions $F_\kappa(E)$ can be well represented by the $q$-normal $f_{qN}(x|q)$ form for $\kappa$ states at the center of the $E_\kappa$ spectrum and for all $k$ values in $V(k)$. For these $\kappa$, the strength functions are symmetrical in $E$ as is the result with $f_{qN}(x|q)$. In the same situation, it is seen in another set of numerical calculations that the conditional distribution $f_{CqN}$ of $f_{biv-qN}$ also gives a good description of the numerical results \cite{PLA-21}. Most significantly, it is seen in some very early calculations with $k=2$ that the strength functions become asymmetrical in $E$ as $\l|E_\kappa\r|$ increases (towards the spectrum edges) \cite{CK} and this is confirmed more recently for all $k$ \cite{PLA-21}. This property can not be generated by $f_{qN}(x|q)$. From the above, it follows that in general for constructing strength functions we need the knowledge of $\rho(E_\kappa , E)$ or that of the conditional $f_{CqN}$ of this bivariate distribution; note that $\rho(E_\kappa , E)$ is the joint distribution in $E_\kappa$ and $E$ where $E_\kappa$ are the eigenvalues of $H_0$ and $E$ are $H$ 
eigenvalues (see Eq. (6) ahead). It is important to mention here that the conditional $f_{CqN}$ generates the asymmetry mentioned above. We will show, by deriving analytical formulas for the lowest four moments of the strength functions, that indeed $f_{CqN}$ used in \cite{PLA-21} to a good approximation represents strength functions.

Given a set of basis states $\kappa$ generated by a unperturbed $t$-body Hamiltonian $H_0(t)$ in $m$ particle spaces, the system Hamiltonian is $H = H_0(t) + V($k$)$ where $V(k)$ is a $k$-body interaction. Let us say that the eigenstate energies are $E$ and the basis states energies, defined by $H_0$, are $E_\kappa$. Now, the strength function $F_\kappa(E)$ is the conditional density of a bivariate density $\rho(E_\kappa,E)$ \cite{KS,Ko-01}. The strength functions $F_\kappa(E)$ determine completely the wavefunction structure in terms of the $\kappa$ states.
Thus, we can infer about the form of the strength functions and the parameters that define them, provided we can determine $\rho(E_\kappa,E)$, its marginals and conditionals. We show that the strength functions are well represented by conditional $q$-normal distributions and derive the necessary parameters for the same. We also write down the formulas for NPC and $S^{info}$ in terms of strength functions. Now we will give a preview.

Section 2 defines the embedded ensembles, strength functions and its moments along with $q$-normal, bivariate $q$-normal and conditional $q$-normal distributions. The formulas for lowest four moments of conditional $q$-normal distributions are derived in Section 3. Section 4.1 gives lowest four moments of bivariate distribution $\rho(E_\kappa,E)$. The lowest four moments of strength functions are derived in Section 4.2 that are valid for $N \to \infty $ and sufficiently large value for $\lambda$. For completeness, finite $N$ results for parameters are given in Section 4.3. Numerical results and discussion of formulas derived in Sections 3 and 4 are given in Section 5. Finally, Section 6 gives conclusions and future outlook including their possible applications.

\section{Preliminaries}

\subsection{The Model}

Constituents of finite many-body quantum systems such as nuclei, atoms, molecules, small metallic grains, quantum dots, arrays of ultracold atoms, and so on, interact via few-body (mainly two-body) interactions \cite{MF, Fr-71, Bo-71, Br-81, BW, BW-rev, Small, Ko-book}. As is well-known, the classical random matrix ensembles [Gaussian Orthogonal Ensembles (GOE)] incorporate many-body interactions. Embedded ensembles [Embedded Gaussian Orthogonal Ensembles (EGOE)] take into account the few-body nature of interactions and hence, they are more appropriate for analyzing various statistical properties of finite quantum systems \cite{MF, Fr-71, Bo-71, Br-81, BW, Ko-book}.

Given a system of $m$ fermions distributed in $N$ levels interacting via $k$-body $(1 \leq k \leq m)$ interactions, embedded ensembles are generated by representing the few fermion ($k$) Hamiltonian by a classical GOE and then the many-fermion Hamiltonian ($m>k$) is generated by the Hilbert space geometry. In other words, $k$-fermion Hamiltonian is embedded in the $m$-fermion Hamiltonian in the sense that the non-zero $m$-fermion Hamiltonian matrix elements are appropriate linear combinations of the $k$-fermion matrix elements. Due to the $k$-body selection rules, many matrix elements of the $m$-fermion Hamiltonian will be zero unlike in a GOE.

The random $k$-body Hamiltonian in second quantized form for a EGOE$(k)$ is,
\begin{equation}
V(k) = \displaystyle\sum_{\tau,\;\gamma} \; v^{\tau,\gamma}_{k} \; \psi^\dagger(k; \tau) \; \psi(k;\gamma) \;.
\label{eq-1}
\end{equation} 
Here, $\tau$ and $\gamma$ are $k$-particle configuration states in occupation number basis. Distributing $k$ fermions in agreement with Pauli's exclusion principle in $N$ single particle (sp) states will generate the complete set of these distinct configurations. Total number of these configurations are ${{N}\choose{k}}$. In occupation number basis, we order the sp levels (denoted by $\mu_i$) in increasing order, $\mu_1 \leq \mu_2 \leq \cdots \leq \mu_N$. Operators $\psi^\dagger(k; \tau)$ and $\psi(k;\gamma)$ respectively are $k$-particle creation and annihilation operators for fermions, i.e. $\psi^\dagger(k; \tau) = \prod_{i=1}^{k} a^\dagger_{\mu_i}$ and $\psi(k;\gamma) = \prod_{i=1}^{k} a_{\mu_i}$. The sum in Eq. \eqref{eq-1} stands for summing over a subset of $k$-particle creation and annihilation operators. These $k$-particle operators obey the usual anti-commutation relations for fermions. 

In Equation \eqref{eq-1}, $v^{\tau,\;\gamma}_{k}$ is chosen to be a $\binom{N}{k}$  dimensional GOE in $k$-fermion spaces. That means $v^{\tau,\;\gamma}_{k}$ are anti-symmetrized few-body matrix elements for fermions chosen to be randomly distributed independent Gaussian variables with zero mean and variance
\begin{equation}
{\overline{v^{\tau,\gamma}_{k} \; v^{\tau^\prime,\gamma^\prime}_{k}}} = v^2 \; \left( {\delta_{\tau,\gamma^\prime}} {\delta_{\tau^\prime,\gamma}} + {\delta_{\tau,\tau^\prime}} {\delta_{\gamma^\prime,\gamma}} \right) \;.
\label{eq-2}
\end{equation} 
Here, the bar denotes ensemble averaging and we choose $v=1$ without loss of generality. 

Distributing the $m$ fermions in all possible ways in $N$ levels generates the many-particle basis states defining $d = \binom{N}{m}$ dimensional Hilbert space. The action of the Hamiltonian operator $V(k)$ defined by Equation \eqref{eq-1} on the many-fermion states generates the EGOE($k$) ensemble in $m$-fermion spaces.

\subsection{Strength functions}
   
Let us begin with a finite quantum many-particle system with $m$ fermions in $N$ sp states defined by the Hamiltonian, 
\be
H = H_0(t) + \lambda V(k)
\label{eq.hh}
\ee
where $H_0$ is a $t$-body operator, $V$ is a $k$-body operator and $\lambda$ is the strength parameter. We will assume that $t < k$ and for $m$ fermions, obviously interaction rank $k \leq m$. In many physical applications $t=1$ with $H_0$ representing a mean-field one-body Hamiltonian \cite{Ko-book, Alhassid-review, Pap-review, Gu-review, Ma-14}. 

Our purpose is to study the structure of eigenfunctions of $H$ expanded in terms of the unperturbed $H_0$ eigenstates (basis states). Denoting $\l|\kappa, \alpha\ran$ as the eigenstates of $H_0$ forming a complete set with $H_0 \l|\kappa, \alpha\ran = E_\kappa \l|\kappa, \alpha\ran$ and $\l|E, \beta\ran$ as the eigenstates of $H$ forming a complete set with $H\l|E, \beta \ran = E \l|E, \beta \ran$ (with $\alpha$ and $\beta$ labeling the respective degeneracies in $H_0$ and $H$ spectrums), we can expand the eigenstates of $H_0$ in the eigenbasis of $H$ as 
\begin{equation}
\l|\kappa, \alpha\ran = \sum_{E, \beta} C_{\kappa, \alpha}^{E, \beta} \l|E, \beta \ran \;.
\label{eq.ovlp}
\end{equation}
Here, $C_{\kappa, \alpha}^{E, \beta} = \lan E, \beta |\kappa, \alpha \ran$ are expansion coefficients of a $\l|\kappa, \alpha\ran$ state in terms of the $\l|E, \beta \ran$ states. Dimension $d$ gives number of $\l|E, \beta \ran$ states and also $\l|\kappa, \alpha\ran$ states for a $m$ fermion system.

Strength function $F_\kappa(E)$ for a $\l|\kappa, \alpha\ran$ state gives the intensity with which a $\l|E, \beta \ran$ state is contained in the $\l|\kappa\ran$ state. Then, with $\lan ---\ran$ denoting average and $\lan\lan --- \ran\ran$ denoting trace, $F_\kappa(E)$ is given by
\be
\barr{rcl}
F_\kappa(E)  & = & \lan \delta(H-E)\ran^{\kappa} = \dis\f{1}{d \cdot \rho_1(E_\kappa)}\;
\lan\lan \delta(H-E) \ran\ran^\kappa \\
& = & \dis\f{1}{d \cdot \rho_1(E_\kappa)}\; \dis\sum_{\alpha \in \kappa;\,\kappa} \lan \kappa , \alpha \mid \delta(H-E) \mid \kappa , \alpha\ran \\
& = & \dis\f{1}{d \cdot \rho_1(E_\kappa)} \dis\sum_{\alpha \in \kappa , \beta \in E;\,\kappa\,,E}
\l|C_{\kappa , \alpha}^{E , \beta}\r|^2 = 
\overline{\l| C_{E_\kappa}^E \r|^2}\; \l[d \cdot \rho_2(E)\r]\;.
\earr \label{eq.fke}
\ee
Here, $d \cdot \rho_1(E_\kappa)$ gives number of $H_0$ states with same basis state energy $E_\kappa$ and similarly $d \cdot \rho_2(E)$ gives number of eigenstates of $H$ with same eigen energy $E$. We use the notation $d \cdot \rho_-(-)$ as $d\rho_-(-)$ may be considered as a differential. Thus, Eq. (\ref{eq.fke}) takes into account degeneracies in the $E_\kappa$ and $E$ spectra and $\overline{\l|C_{E_\kappa}^{E}\r|^2}$ is average of $\l|C_{\kappa, \alpha}^{E, \beta}\r|^2$ taken over the degenerate $\kappa$ states and $E$ states. 

Using Eq. \eqref{eq.fke}, it is easy to see that $F_\kappa(E)$, which is a function of eigen energies $E$ with fixed basis state energy $E_\kappa$, is a conditional density of a bivariate distribution in $E_\kappa$ and $E$ defined by \cite{KS,Ko-01},
\be
\rho(E_\kappa,E) = (1/d) \lan\lan \delta(H_0 - E_\kappa) \delta(H-E)\ran\ran^m = d\; \overline{\l| C_{E_\kappa}^E \r|^2} \rho_1(E_\kappa)\,\rho_2(E)\;.
\label{eq.fke1}
\ee
The $\rho_2(E) = \lan \delta(H-E)\ran^m$ is the eigenvalue density generated by $H$ and similarly, $\rho_1(E_\kappa) = \lan \delta(H_0 - E_\kappa) \ran^m$ is the eigenvalue density generated by $H_0$. Note that $\rho_1(E_\kappa)$ and $\rho_2(E)$ are the marginals of $\rho(E_\kappa ,E)$ and all the $\rho$'s are normalized to unity. With these, we have the important relation \cite{KS,Ko-01}
\be
F_\kappa(E) = \dis\f{\rho(E_\kappa,E)}{\rho_1(E_\kappa)}\;.
\label{eq.fke2}
\ee
As we will be using moments method (in the moment method, one evalutes the lower order moments of a distribution function to infer the distribution [2, 27, 28]) for deriving the distributions of interest, let us mention that the $P$-th order moments $M_P$ of $\rho_1(E_\kappa)$ and $\rho_2(E)$ are $\lan H_0^P\ran^m$ and $\lan H^P \ran^m$ respectively. Note that $P = 1$ defines centroid $\epsilon_1 = \lan H_0 \ran^m$ and $P = 2$ defines variance $\sigma_1^2 = \lan H_0^2\ran^m - (\lan H_0 \ran^m)^2$ of $\rho_1(E_\kappa)$. Similarly, $\epsilon_2 = \lan H \ran^m$ and $\sigma_2^2 = \lan H^2\ran^m - (\lan H \ran^m)^2$ respectively define the centroid and variance of $\rho_2(E)$. The bivariate moments $M_{PQ}$ of $\rho(E_\kappa ,E)$ are
\be
\barr{rcl}
M_{PQ} & = & \lan H_0^P H^Q\ran^m \\
& = & d^{-1} \dis\sum_{\alpha \in \kappa\,,\,\beta \in \kappa^\pr; \kappa\,,\kappa^\pr} \lan \kappa, \alpha \mid H_0^P \mid \kappa^\pr, \beta\ran \lan \kappa^\pr, \beta \mid H^Q \mid \kappa, \alpha \ran \\
& = & d^{-1} \dis\sum_{\alpha \in \kappa\,,\,\beta \in \kappa^\pr; \kappa\,,\kappa^\pr} \lan \kappa, \alpha \mid H_0^P \mid \kappa^\pr, \beta\ran \lan \kappa^\pr, \beta \mid H^Q \mid \kappa, \alpha \ran \delta_{\kappa , \kappa^\pr}\, \delta_{\alpha , \beta}\\
& = & d^{-1} \dis\sum_{\alpha \in \kappa; \kappa} E_\kappa^P \; \lan \kappa\,, \alpha \mid H^Q \mid \kappa\,, \alpha\ran \\
& = & d^{-1}  \dis\sum_{\alpha \in \kappa,\;\beta \in E,\;\beta^\pr \in E^\pr; \kappa\,,E\,,E^\pr} C_{\kappa, \alpha}^{E, \beta} \;C_{\kappa, \alpha}^{E^\pr , \beta^{\pr}}\; E_\kappa^P\, \lan E, \beta \mid H^Q \mid E^\pr , \beta^\pr \ran \\
& = & d^{-1} \dis\sum_{\alpha \in \kappa,\;\beta \in E; \kappa\,,E} {\l|C_{\kappa, \alpha}^{E, \beta}\r|}^2\; E_\kappa^P E^Q \\
& = & d \dis\sum_{E_\kappa, E} E_\kappa^P E^Q \;\overline{\l|C_{E_\kappa}^E\r|^2}\;
\rho_1(E_\kappa)\,\rho_2(E) \;.
\earr \label{eq.moment}
\ee
In the first step in Eq. (\ref{eq.moment}), we have expanded in terms of basis states $\l|\kappa, \alpha\ran$ using the property that traces are invariant under unitary transformations. Given the moments $M_{PQ}$, the central moments $\cam_{PQ}$ follow from Eq. (\ref{eq.moment}) by replacing $E_\kappa$ by $E_\kappa -\epsilon_1$ and $E$ by $E - \epsilon_2$ and the reduced moments, free of location and scale, are $\mu_{PQ} = (\cam_{PQ})/(\sigma_1^P \sigma_2^Q)$ respectively.

\subsection{Conditional $q$-normal distribution}

Let us begin with the $q$-normal distribution $f_{qN}(x|q)$ \cite{Ismail,Sza-1}, with $x$ being a standardized variable (then $x$ is zero centered with variance unity),
\be
f_{qN}(x|q) = \dis\frac{\dis\sqrt{1-q} \dis\prod_{k^\pr=0}^{\infty} \l(1-
q^{k^\pr +1}\r)}{2\pi\,\dis\sqrt{4-(1-q)x^2}}\; \dis\prod_{k^\pr=0}^{\infty}
\l[(1+q^{k^\pr})^2 - (1-q) q^{k^\pr} x^2\r]\;.
\label{eq.qbiv-1}
\ee
The $f_{qN}(x|q)$ is defined over $S(q)$ with
$$
S(q) = \l(-\dis\frac{2}{\dis\sqrt{1-q}}\;,\;+\dis\frac{2}{\dis\sqrt{1-q}}\r) \;.
$$
In this paper, we consider $0 \leq q \leq 1$. Note that the integral of $f_{qN}(x|q)$ over $S(q)$ is unity. For $q=1$ taking the limit properly will give $S(q)=(-\infty , \infty)$. It is easy to see that $f_{qN}(x|1)=(1/\sqrt{2\pi}) \, \exp (-x^2/2)$, the Gaussian and $f_{qN}(x|0)=(1/2\pi) \sqrt{4-x^2}$, the semi-circle.

Going further, bivariate $q$-normal distribution $f_{biv-qN}(x,y|\xi , q)$ as given in \cite{Sza-1}, with $x$ and $y$ standardized variables, is defined as follows,
\be
\barr{l}
f_{biv-qN}(x,y|\xi , q) = f_{qN}(x|q) \; f_{qN}(y|q) \; h(x,y|\xi , q)\;;\\ \\
h(x,y|\xi ,q)= \dis\prod_{k^\pr=0}^\infty \dis\frac{1-\xi^2 q^{k^\pr}}{
(1-\xi^2 q^{2k^\pr})^2 -(1-q)\,\xi\, q^{k^\pr}\,(1+\xi^2 q^{2k^\pr})\,xy +
(1-q)\xi^2 q^{2k^\pr} (x^2 +y^2)}\;,
\earr \label{eq.qbiv-2}
\ee
where $\xi$ is the bivariate correlation coefficient. The conditional $q$-normal
densities $f_{CqN}$ are then,
\be
\barr{l}
f_{biv-qN}(x,y|\xi , q) = f_{qN}(x|q) \; f_{CqN}(y|x; \xi , q) =
f_{qN}(y|q) \; f_{CqN}(x|y; \xi , q)\;;\\ \\
f_{CqN}(x|y; \xi , q)=f_{qN}(x|q) \; h(x,y|\xi ,q)\;,\\ \\
f_{CqN}(y|x; \xi , q)=f_{qN}(y|q) \; h(x,y|\xi ,q)\;.
\earr \label{eq.qbiv-3}
\ee
A very important property of $f_{CqN}$ is
\be
\dis\int_{S(q)} He_n(x|q) \; f_{CqN}(x|y; \xi ,q) \; dx = \xi^n \; He_n(y|q)\;.
\label{eq.qbiv-4}
\ee
Here, $He_n$ are Hermite polynomials. With $q$-numbers $\l[n\r]_q = [1-q^n]/[1-q] = 1+q +q^2 + \ldots + q^{n-1}$ (note that $\l[0\r]_q=0$), the $q$-Hermite polynomials are defined by the relation
\be
\barr{l}
He_{n+1}(x|q) = x \,He_n(x|q) - \l[n\r]_q \,He_{n-1}(x|q)\;\;\mbox{with}\;\; 
n \geq 1; \\
He_{-1}(x|q)=0,\;\;He_0(x|q)=1\;.
\earr
\label{eq.qnew1}
\ee
Note that $He_n(x|1)=He_n(x)$, the Hermite polynomials with respect to $1/\sqrt{2\pi}\,\exp(-x^2/2)$. Also, $He_n(x|0)=U_n(x/2)$, the Chebyshev polynomials. Putting $n=0$ in Eq. (\ref{eq.qbiv-4}), it can be verified that $f_{CqN}$ and hence $f_{biv-qN}$ are normalized to unity over $S(q)$. We will make use of Eq. (\ref{eq.qbiv-4}) to derive the lowest four moments of $f_{CqN}$. A general formula, though complicated, valid for moments of any order is given in \cite{Szb-2}. For $q=1$, $f_{CqN}$ reduces to the conditional Gaussian of a bivariate Gaussian and hence in the $q=1$ limit, the skewness $\gamma_1$ and excess $\gamma_2$ of $f_{CqN}$ are zero.

\section{Formulas for the lowest four moments of $f_{CqN}$}

In this Section, we will derive the lowest four moments of conditional $q$-normal distribution $f_{CqN}(x|y;\xi , q)$ defined by Eq. \eqref{eq.qbiv-3}. It is easy to see that the first moment is given by
\be
M_1=\dis\int_{S(q)} x f_{CqN}(x|y) dx = \xi He_1(y) = \xi y\;.
\label{eq.qbiv-5}
\ee
Here, $He_1$ is the first-order Hermite polynomial. Now, the central moments $\mu_{r}$ of $f_{CqN}$ are defined by
\be
\cam_{r}(y) = \dis\int_{S(q)} (x -\xi y)^r\,f_{CqN}(x|y;\xi ,q)\, dx\;.
\label{eq.qbiv-6}
\ee
Note that
\be
\barr{rcl}
x & = & He_1\;,\;x^2 =  He_2+He_0\;,x^3 = He_3 + (2+q)He_1\;,\\
x^4 & = & He_4 + (3+2q+q^2)He_2 +(2+q)He_0\;.
\earr \label{eq.qbiv7}
\ee
Recall that $He_n$ stands for Hermite polynomials $He_n(x|q)$ and $He_0(x|q)=1$. Now, $\cam_2=\sigma^2$ is,
\be
\barr{rcl}
\mu_2(y) & = & \sigma^2(y) = \dis \int_{S(q)} (x-\xi y)^2 f_{CqN}(x|y) dx \\
& = & \dis \int_{S(q)} [He_2(x)+He_0(x)] \, f_{CqN}(x|y) dx -\xi^2 y^2 \\ \\
& = & \xi^2 He_2(y) + 1 -\xi^2 y^2 = 1-\xi^2 \;.
\earr \label{eq.qbiv8}
\ee
In deriving Eq. \eqref{eq.qbiv8}, first we wrote $x^2$ and $x$ in terms of $He_n(x)$ using Eq. (\ref{eq.qbiv7}) and then used Eq. (\ref{eq.qbiv-4}). Finally, Eq. (\ref{eq.qbiv7}) is used again to write $He_n(y)$ in terms of $y^r$. Going further, the third moment is
\be
\cam_3(y) = \dis \int_{S(q)} (x-\xi y)^3 f_{CqN}(x|y) dx = -(1-\xi^2) (1-q)(\xi y)\;.
\label{eq.qbiv9}
\ee
Note that, we first expanded $(x-\xi y)^3$, changed $x^r$ into $He_n(x)$ using Eq. (\ref{eq.qbiv7}) and then applied Eq. (\ref{eq.qbiv-4}) for evaluating the integrals. Finally, changed the $He_n(y)$ into $y^r$. Now, the reduced third moment $\mu_3(y)$ = $\gamma_1(y)=\cam_3(y)/\sigma^3(y)$ is given by (as $\sigma^2(y) = (1-\xi^2)$, independent of $y$),
\be
\mu_3(y) = \gamma_1(y) = -\dis\frac{\xi (1-q)\,y}{\dis\sqrt{1-\xi^2}}\;.
\label{eq.qbiv10}
\ee
Note that $\gamma_1$ is the skewness (or asymmetry) parameter. Proceeding similarly, we have for the fourth central moment
\be
\mu_4(y) = \dis\frac{1}{(1-\xi^2)^2}\;\dis\int_{S(q)} (x-\xi y)^4 f_{CqN}(x|y) dx = (2+q) +
\dis\frac{(1-q)^2 \xi^2 y^2 + \xi^2 (1- q^2)}{(1-\xi^2)}\;.
\label{eq.qbiv11}
\ee
Then, $\gamma_2(y)=\mu_4(y) - 3$ is
\be
\gamma_2(y) = (q-1) +
\dis\frac{(1-q)^2 \xi^2 y^2 + \xi^2 (1-q^2)}{(1-\xi^2)}\;.
\label{eq.qbiv12}
\ee
Note that $\gamma_2$ is the excess parameter.
It is easy to see from Eqs. (\ref{eq.qbiv10}) and (\ref{eq.qbiv12}) that $\gamma_1=0$ and $\gamma_2=0$ for $q=1$ correctly as required for a Gaussian. Note that $f_{CqN}$ has the important property that it is, in general, an asymmetrical function in its variable. The formulas given here for the first four moments when applied to $F_\kappa(E)$ will test if the conditional $q$-normal is a good representation or not. Now, we will derive formulas for first four moments of $\rho(E_\kappa, E)$ and $F_\kappa(E)$ defined in Section 2.2 using the random matrix model adopted in Section 2.1.

\section{Binary correlation results}

Using random matrix description of $H_0$ and $V$ operators, defined by Eq. \eqref{eq-1}, ensemble averaged moments can be evaluated for Hamiltonian $H$ defined by Eq. \eqref{eq.hh}, in the `dilute limit' for fermions defined by $N \rightarrow \infty$, $m \to \infty$, $m/N \to 0$, $k << m$, $t << m$, using the so called binary correlation approximation (BCA). This approximation allows one to derive averages (or traces) involving arbitrary products of creation and annihilation operators by reducing it to sums of products of pairs of these operators. This removes the dependence of the moments on the number of sp states $N$; see \cite{MF,Br-81,Ko-book,steve} for further details of BCA. In the end of this Section, we will give some useful finite $N$ formulas. 

\subsection{Lower order bivariate moments of $\rho(E_\kappa, E)$}

One approach to derive the form of $F_\kappa(E)$ is to use Eq. (\ref{eq.fke2}) by constructing $\rho(E_\kappa,E)$. From the results in \cite{MK}, clearly $\rho_1(E_\kappa)$ will be a $q$-normal distribution defined in Eq. (\ref{eq.qbiv-1}). Then, it is natural to examine if $\rho(E_\kappa, E)$ follows bivariate $q$-normal form given by Eq. (\ref{eq.qbiv-2}). Now, we will study the appropriateness of representing strength functions by bivariate $q$-normal distributions by deriving formulas for the lower order bivariate moments of $\rho(E_\kappa,E)$.  

In order to evaluate the lower order moments of $\rho(E_\kappa, E)$, we will consider a random matrix representation of the $H$ operator. Towards this end, we will represent $H_0(t)$ by EGOE($t$) and $V(k)$ by EGOE($k$), defined by Eq. \eqref{eq-1}. In addition, we assume that the EGOE($t$) and EGOE($k$) are independent. With these, using the Hamiltonian operator given by Eq. (\ref{eq.hh}) for each member of the ensemble, the $m$-fermion $H$ matrix can be constructed and so also the $H_0(t)$ and $V(k)$ matrices in $m$-fermion spaces. These will give the bivariate moments $M_{PQ} = \lan H_0^P H^Q\ran^m$ generated by each member of the ensemble. Averaging over the ensemble will then give $\overline{M_{PQ}}$; with `bar' denoting ensemble average.

Firstly, the independence of the EGOE's for $H_0(t)$ and $V(k)$ operators implies 
\be
\overline{\lan \l[H_0(t)\r]^r \l
[V(k)\r ]^s\ran^m} = \overline{\lan \l[ H_0(t)\r]^r\ran^m} \;\;\overline{\lan \l[V(k)\r]^s\ran^m}\;.
\label{eq.mpq1}
\ee
Also, EGOE representation gives
\be
\barr{l}
\overline{\lan \l[H_0(t)\r]^r\ran^m} = 0\;\;\mbox{for}\;\;r\;\mbox{odd}\;,\;
\overline{\lan \l[V(k)\r]^s\ran^m} = 0\;\;\mbox{for}\;\;s\;\mbox{odd}\;.
\earr \label{eq.mpq2}
\ee
From now on, for brevity, we will drop $t$ and $k$ in $H_0$ and $V$ respectively. Equation (\ref{eq.mpq2}) immediately gives the result that the centroids of $\rho(E_\kappa , E)$ are zero,
\be
\overline{\lan H_0\ran^m} =0\;,\;\;\; \overline{\lan H \ran^m} = 0\;.
\label{eq.mpq3}
\ee
With this, $ \lan H_0^P H^Q\ran^m$ will define the central moments $\cam_{PQ}$. Now, BCA will give the following results for the variances,
\be
\barr{rcl}
\sigma^2_{H_0} & = & \overline{\lan H^2_0\ran^m} =
\dis\binom{m}{t}\,\dis\binom{N}{t}\;, \\ \\
\sigma^2_V & = & \overline{\lan V^2\ran^m} = \dis\binom{m}{k}\,\dis\binom{N}{k}\;, \\ \\
\sigma^2_H = \overline{\lan H^2\ran^m} & = &  \sigma^2_{H_0} + \lambda^2 \; \sigma^2_V  
= \dis\binom{m}{t}\,\dis\binom{N}{t} +
\lambda^2\,\dis\binom{m}{k}\,\dis\binom{N}{k}\;. \\
\earr \label{eq.mpq4}
\ee
Here we have used Eqs. (\ref{eq.mpq1}) and (\ref{eq.mpq2}) that give $\overline{\lan H_0 \, V\ran^m}=0$. 

Going further, we need the reduced moments $\mu_{PQ}$,
\be
\mu_{PQ} = \dis\frac{\overline{\lan H_0^P H^Q\ran^m}}{\sigma_{H_0}^P\,\sigma_H^Q}\;.
\label{eq.mpq5}
\ee
The first reduced moment of interest is the correlation coefficient $\xi$. Using $\overline{\lan H_0 \, V\ran^m}=0$, $\xi$ is given by, 
\be
\barr{rcl}
\xi & = & \dis\frac{\overline{\lan H_0 H\ran^m}}{\sigma_{H_0} \sigma_H} = \dis\frac{\overline{\lan H^2_0 \ran^m}}{\sigma_{H_0} \sigma_H} = \dis\frac{\sigma_{H_0}}{\sigma_H} =\dis\sqrt{\dis\frac{\binom{m}{t}}{\binom{m}{t} + \lambda^2 \; {\binom{N}{t}}^{-1}\,\binom{N}{k}\;\binom{m}{k}}}\;.
\earr \label{eq.mpq6}
\ee

Going to higher order moments ($P+Q \geq 3)$, we have $\mu_{PQ}=0$ for $P+Q$ odd. Thus, the fourth order moments $\mu_{PQ}$ with $P+Q=4$ are most important,
\be
\barr{l}
\mu_{40} = \dis\frac{\overline{\lan \l[H_0\r]^4\ran^m}}{\sigma_{H_0}^4}\;,\;\;\;\mu_{04} = \dis\frac{\overline{\lan \l[H_0 + \lambda V\r]^4\ran^m}}{\sigma_H^4}\;, \\
\\
\mu_{31} = \dis\frac{\overline{\lan \l[H_0\r]^3 (H_0+ \lambda V)\ran^m}}{\sigma_{H_0}^3 \sigma_H}\;,\;\;\;\mu_{13} = \dis\frac{\overline{\lan H_0\,\l[H_0 + \lambda V\r]^3\ran^m}}{\sigma_{H_0} \sigma^3_H}\;, \\
\\
\mu_{22}=\dis\frac{\overline{\lan H_0^2\;\l[H_0+ \lambda V\r]^2\ran^m}}{\sigma_{H_0}^2 \sigma_H^2}\;.
\earr \label{eq.mpq8}
\ee
From now on, we will drop the `bar' over the $m$-fermion averages. Using BCA, 
\be
\mu_{40}  =  \l[\lan H_0^2(t)\ran^m\r]^{-2}\;\lan H_0^4(t)\ran^m = 2 + q^{h}\;;\;\;\; q^{h} = \binom{m}{t}^{-1}\,\binom{m-t}{t} \;.
\label{eq.mpq9}
\ee
Similarly, introducing $\hat{V}=V/\sigma_V$ and $\hat{H}_0 = H_0/\sigma_{H_0}$, we have
\be
\barr{rcl}
\mu_{04} & = & \dis\frac{\lan H^4\ran^m}{\sigma_H^4} = \dis\frac{\lan (H_0 + \lambda V)^4\ran^m}{\sigma_H^4} \\ \\
& = & \dis\frac{\lan H_0^4\ran^m}{\sigma_H^4} + 4\lambda^2 \dis\frac{\lan H_0^2\ran^m \lan V^2 \ran^m}{\sigma_H^4} + \lambda^4 \dis\frac{\lan V^4\ran^m}{\sigma_H^4} + 2\lambda^2 \dis\f{\lan H_0 V H_0 V\ran^m}{\sigma_H^4} \\ \\
& = & \xi^4\l(2+q^{h}\r) + 4 \xi^2 (1-\xi^2) + 2 \xi^2 (1-\xi^2) (q^{hv}) +
(1-\xi^2)^2 \l(2+ q^v\r) \\
& = & 2 + q^H\;;\\
q^v & = & \dis\frac{\binom{m-k}{k}}{\binom{m}{k}}\;,\;\;\;q^{hv}=\lan \hat{H}_0 \hat{V} \hat{H}_0 \hat{V}\ran^m = \dis\frac{\binom{m-t}{k}}{\binom{m}{k}}\;,\\ \\
q^H & = & \l[\xi^4 q^{h} + (1-\xi^2)^2 q^v + 2 \xi^2 (1-\xi^2) 
q^{hv} \r]\;.\\
\earr \label{eq.mpq10}
\ee
Here we have used Eqs. (\ref{eq.mpq1}) and (\ref{eq.mpq2}) in simplifications.  

Using $\lan H_0^3 H\ran^m = \lan H_0^4\ran^m$, we have 
\be
\barr{rcl}
\mu_{31} & = & \dis\f{\lan H_0^4\ran^m}{\sigma^3_{H_0} \sigma_H} = \dis\f{\sigma_{H_0}}{\sigma_H} \l(2+q^{h}\r) \\
& = & \xi\,\mu_{40} = \dis\f{2\binom{m}{t} + \binom{m-t}{t}}{\sqrt{\binom{m}{t} \l[\binom{m}{t} + \lambda^2 \; {\binom{N}{t}}^{-1}\,\binom{N}{k} \; \binom{m}{k}\r]}}\;.
\earr \label{eq.mpq11}
\ee
Note that $q^{h}$ is given by Eq. (\ref{eq.mpq9}). Similarly, using ${\lan H_0 H^3\ran} = {\lan H_0^4 \ran} + 2 \lambda^2 {\lan H_0^2\ran}\;{\lan V^2\ran}$ + $\lambda^2 {\lan H_0 V H_0 V\ran}$ gives,
\be
\barr{rcl}
\mu_{13} & = & \xi \l[2 + \xi^2 q^{h} + (1-\xi^2) q^{hv} \r] \\
& = & \xi \l[2 + \dis\f{\binom{m-t}{t} + \lambda^2 \; {\binom{N}{t}}^{-1}\,\binom{N}{k} \; \binom{m-t}{k}}{\binom{m}{t} + \lambda^2 \; {\binom{N}{t}}^{-1}\,\binom{N}{k} \; \binom{m}{k}}\r]\;.
\earr \label{eq.mpq12}
\ee
Finally, using $\overline{\lan H_0^2 H^2\ran} = \overline{\lan H_0^4\ran} + \lambda^2 \overline{\lan H_0^2\ran}\;\overline{\lan V^2\ran}$,
\be
\barr{rcl}
\mu_{22} & = & \xi^2 \l(2+ q^{h}\r) + (1-\xi^2) \\
& = & 1 + \dis\frac{\binom{m}{t} + \binom{m-t}{t}}{\binom{m}{t} + \lambda^2 \; {\binom{N}{t}}^{-1}\,\binom{N}{k} \; \,\binom{m}{k}} \;. \\
\earr \label{eq.mpq13}
\ee
Formulas in Eqs. \eqref{eq.mpq9}-\eqref{eq.mpq12} show that in general $\mu_{PQ} \neq \mu_{QP}$. Therefore, as $k$ increases towards $m$, $\rho(E_\kappa ,E)$ will not be in general well represented by $f_{biv-qN}$ as this demands $\mu_{PQ}=\mu_{QP}$ for all $P$ and $Q$. This result also implies that $F_\kappa(E)$ will be asymmetrical in $E$. Although the use of Eq. (\ref{eq.fke2}) with $f_{biv-qN}$ for $\rho(E_\kappa ,E)$ is ruled out, it will not preclude the possibility of representing $F_{\kappa}(E)$ directly as a conditional $q$-normal  distribution $f_{CqN}$ with its parameters appropriately defined.

\subsection{First four moments of strength functions}

In order to establish that the strength functions $F_\kappa(\hat{E})$ follow conditional $q$-normal densities $f_{CqN}(\hat{E}|\ek ; \xi , q^{hv})$,  we will derive formulas for the first four moments of the strength functions $F_\kappa(E)$ generated by $H$ defined by Eq. \eqref{eq.hh}. The strength functions are defined for each $E_\kappa$ energies that are $H_0$ eigenvalues.
Again, we will represent $H_0(t)$ and $V(k)$ in Eq. \eqref{eq.hh} by independent EGOE($t$) and EGOE($k)$ ensembles respectively and use BCA to derive formulas for the moments. Scaling the eigenvalues $E$ with their width $\sigma_H$, the moments of $F_\kappa(E)$ are given by
\be
M_r(E_\kappa) = \dis\f{\lan H^r \ran^{\kappa}}{\l(\sigma_H\r)^r}\;.
\label{eq.fek1} 
\ee
Although it is not shown explicitly in Eq. (\ref{eq.fek1}), we are considering ensemble averaged $M_r$. It is important to note: (i) $\lan H_0^p \ran^{\kappa} = E^p_\kappa$ as $\kappa$ are eigenstates of $H_0$ with eigenvalues $E_\kappa$; (ii) we need expectation values of operators for evaluating $M_r$. Given an operator $K$, expectation value $\lan K \ran^{\kappa}$ follows for example from a polynomial expansion \cite{DFW,KH-10},
\be
\barr{l}
\lan K \ran^{\kappa} = \dis\sum_\mu \lan K P_\mu(H_0) \ran^m P_\mu(E_\kappa)\;;\\
P_0(x) = 1,\;\;\;P_1(x) = \hat{x},\;\;\;P_2(x) = \dis\frac{(\hat{x})^2-1}{\dis\sqrt{\mu_4-1}}\;.
\earr \label{eq.fek2}
\ee
Note that we are using zero centered $x$ and $\hat{x}=x/\sigma$ where $\sigma$ is the width of the variable $x$ and similarly $\mu_4$ is its fourth reduced moment. We assume that the third reduced moment of $x$ is zero as is the situation with $E$ and $E_\kappa$ when we use EGOE($t$) and EGOE($k$) ensembles; the energies are also zero centered. The expansion in Eq. (\ref{eq.fek2}) converges in general and therefore often only first two or three terms in the sum suffice \cite{DFW}; see \cite{DFW,KH-10} for the general definition of the polynomials $P_\mu$. In evaluating $M_r(E_\kappa)$, we will often use the result, as the $H_0$ and $V$ ensembles are independent,
\be
\lan V^r \ran^{\kappa} = \dis\sum_\mu \lan V^r P_\mu(H_0) \ran^m P_\mu(E_\kappa) = \lan V^r\ran^m + \dis\sum_{\mu \neq 0} \lan V^r \ran^m \lan P_\mu(H_0)\ran^m P_\mu(E_\kappa) = \lan V^r \ran^m \;.
\label{eq.fek3}
\ee
Note that by definition $\lan P_\mu(H_0)\ran^m=0$ for $\mu \neq 0$. In addition, we also have 
\be
\lan V^r \ran^m =0 \;\;\mbox{for}\;\;r\;\;\;\mbox{odd}
\label{eq.fek4}
\ee
and it is non-zero for $r$ even. Finally, we will use the following relations to convert the moments $M_r$ into central moments $\cam_r$,
\be
\barr{rcl}
\cam_2 & = & M_2 - M_1^2\;,\\
\cam_3 & = & M_3 - 3M_2 M_1 +2M_1^3 \;,\\
\cam_4 & = & M_4 - 4M_3 M_1 +6M_2 M_1^2 -3M_1^4\;.
\earr \label{eq.fek5}
\ee
 
Using Eq. (\ref{eq.fek4}), the centroid $M_1(E_\kappa)$ is
\be
M_1(E_\kappa) = \dis\f{\lan H \ran^{\kappa}}{\sigma_H} = \dis\frac{\lan H_0\ran^{\kappa} + \lambda \lan V \ran^{\kappa}}{\sigma_H} = \xi\, \ek\;.
\label{eq.fek6}
\ee
It is important to note that $\xi \ek = E_\kappa/\sigma_H$. Now, the second moment $M_2(E_\kappa)$ is,
\be
\barr{rcl}
M_2(E_\kappa) & = & \dis\f{\lan H^2 \ran^{\kappa}}{\sigma^2_H} = \dis\f{\lan H^2_0 +\lambda^2 V^2 + \lambda(H_0V + VH_0) \ran^{\kappa}}{\sigma^2_H} \\ \\
& = & \dis\f{E^2_\kappa}{\sigma_H^2} + \lambda^2 \dis\f{\lan V^2\ran^m}{\sigma^2_H} + 2 \lambda \dis\f{E_\kappa \lan V \ran^m}{\sigma^2_H}
= \xi^2 (\ek)^2 + (1-\xi^2) \\ \\
\Rightarrow \cam_2(E_\kappa) & = & (1-\xi^2) \;.
\earr \label{eq.fek7}
\ee
Here we have used Eqs. (\ref{eq.fek3}), (\ref{eq.fek4}), (\ref{eq.fek5}) and (\ref{eq.fek6}) in simplifications. As seen from Eq. (\ref{eq.fek7}), the variance $\cam_2(E_\kappa)$ is independent of $E_\kappa$. Now let us consider $M_3(E_\kappa)$,
\be
\barr{rcl}
M_3(E_\kappa) & = & \dis\f{\lan H^3 \ran^{\kappa}}{\sigma^3_H} = \dis\f{ \lan \l( H^2_0 +\lambda^2 V^2 + \lambda (H_0V + VH_0)\r)\l(H_0 +\lambda V\r) \ran^{\kappa}}{\sigma^3_H} \\ \\
& = & \dis\f{ \lan H_0^3 + \lambda^3 V^3 + \lambda^2 (H_0 V^2 + V^2H_0) + \lambda (H^2_0 V + VH^2_0) + \lambda^2 VH_0V + \lambda H_0 V H_0 \ran^{\kappa}}{\sigma^3_H} \\
& = & \xi^3 (\ek)^3 + 2 \xi (1-\xi^2)\ek + \lambda^2 \dis\f{\lan V H_0 V\ran^{\kappa}}{\sigma^3_H}\;.
\earr \label{eq.fek8}
\ee
Now, the last term $\lan V H_0 V\ran^{\kappa}$ is evaluated using Eq. (\ref{eq.fek2}) by keeping only the first two terms. The first term, in this expansion, with $\lan V H_0 V\ran^m$ will be clearly zero as $H_0$ is an EGOE and the second term gives
\be
\lambda^2 \dis\f{\lan V H_0 V\ran^{\kappa}}{\sigma^3_H} = \lambda^2 \dis\f{\lan V H_0 V H_0\ran^m}{\sigma_{H_0} \sigma^3_H}\;\dis\f{E_\kappa}{\sigma_{H_0}} = \xi (1-\xi^2) \l(q^{hv}\r) (\ek)\;.
\label{eq.fek9}
\ee  
Note that $q^{hv}$ is defined in Eq. (\ref{eq.mpq10}). Now scaling with the variance $\cam_2(E_\kappa)$, will give the following formula for reduced third moment $\mu_3(\ek)$,
\be
\mu_3(\ek) = - \dis\f{\xi \l(1-q^{hv}\r) \ek}{\dis\sqrt{1-\xi^2}} \;.
\label{eq.fek10}
\ee
It is remarkable to note that the first three moments $M_1(E_\kappa)$, $\cam_2(E_\kappa)$ and $\mu_3(E_\kappa)$ as given by Eqs. (\ref{eq.fek6}), (\ref{eq.fek7}) and (\ref{eq.fek10}) respectively are exactly same as the formulas given by $f_{CqN}$; see Eqs. (\ref{eq.qbiv-5}), (\ref{eq.qbiv8}) and (\ref{eq.qbiv10}). The $\xi$ and $q=q^{hv}$ parameters in $f_{CqN}$ are then defined by Eqs. (\ref{eq.mpq6}) and (\ref{eq.mpq10}) respectively. For further verification of this important result and derive any other constraints to be satisfied, let us examine the next fourth reduced moment.

Turning to the fourth moment, firstly we have,
\be
\barr{l}
M_4(E_\kappa) =  \dis\f{\lan H^4 \ran^{\kappa}}{\sigma^4_H} = \dis\f{\lan\l(H^2_0 +\lambda^2 V^2 + \lambda(H_0V + VH_0)\r)^2 \ran^{\kappa}}{\sigma^4_H} \\
= \dis\f{1}{\sigma^4_H} \l[E^4_\kappa + 3 \lambda^2 E^2_\kappa \lan V^2 \ran^{\kappa} + \lambda^4 \lan V^4 \ran^{\kappa} + \lambda^2 \l(\lan H_0 V H_0 V\ran^{\kappa} + \lan V H_0 V H_0\ran^{\kappa} +
\lan V H_0^2 V\ran^{\kappa}\r)\r] \\
= \xi^4 (\ek)^4 + 3 \xi^2 (1-\xi^2) (\ek)^2 + (1-\xi^2)^2 (2+q^v)
+ 2 \xi^2 (1-\xi^2) (\ek) \lan \hat{V} \hat{H}_0 \hat{V}\ran^{\kappa} \\
+ \xi^2 (1-\xi^2) \lan \hat{V} (\hat{H}_0)^2 \hat{V}\ran^{\kappa} \;.
\earr \label{eq.fek11}
\ee
Here we have used the result that $\lan \hat{V}^4 \ran^m = (2+q^v)$; $q^v$ is defined in (\ref{eq.mpq10}). In addition, Eq. (\ref{eq.fek9}) gives 
\be
\lan \hat{V} \hat{H}_0 \hat{V} \ran^{\kappa} = q^{hv}  \hat{E}_\kappa
\label{eq.fek12}
\ee
and similarly Eq. (\ref{eq.fek2}) with first three terms (the second term will be zero) gives,
\be
\barr{l}
\lan \hat{V} (\hat{H}_0)^2 \hat{V}\ran^{\kappa} = \lan \hat{V} (\hat{H}_0)^2 \hat{V}\ran^m + \lan \hat{V} (\hat{H}_0)^2 \hat{V} \l[(\hat{H}_0)^2 -1\r]\ran^m \dis\frac{(\ek)^2 -1}{\mu_{40} -1} \\
= 1 + \dis\frac{(\ek)^2 -1}{\mu_{40} -1} \l[\lan \hat{V} (\hat{H}_0)^2 \hat{V} (\hat{H}_0)^2\ran^m -1\r]\;.
\earr \label{eq.fek13}
\ee
Here, $\mu_{40}$ is given in Eq. \eqref{eq.mpq9}. Now, we can evaluate $\lan \hat{V} (\hat{H}_0)^2 \hat{V} (\hat{H}_0)^2\ran^m$ using BCA. There will be three terms that are evaluated by contracting correlated pairs of $\hat{H}_0$ operators in the first term, contracting the $\hat{H}_0$ operators across $\hat{V}$ operator in the second term and contracting two $\hat{H}_0$ operators across $\hat{H}_0 \, \hat{V}$ (effective rank $t + k$) operator in the third term. Then, we obtain
\be
\lan \hat{V} (\hat{H}_0)^2 \hat{V} (\hat{H}_0)^2\ran^m = 1+ \l(q^{hv}\r)^2 + \l(q^{hv}\r)\;\binom{m}{k}^{-1} \binom{m-k-t}{k} \;.
\label{eq.fek15}
\ee
Now, using the approximation
\be
\mu_{40} = 2 + q^{h} \approx 3
\label{eq.fek14} 
\ee
along with Eqs. (\ref{eq.fek5}), (\ref{eq.fek11}) and \eqref{eq.fek15}, we have 
\be
\barr{rcl}
\mu_4(\ek) & = & \mu_4^0(\ek)\;\l[1 + \Delta(\ek)\r]\;;\\
\mu_4^0(\ek) & = & \l(2+q^{hv}\r) + \dis\f{\xi^2 (\ek)^2 \l(1-q^{hv}\r)^2 + \xi^2 \l[1-\l(q^{hv}\r)^2\r]}{1-\xi^2}\;,\\
\Delta(\ek) & = & \dis\f{\Delta_0(\ek)}{\mu_4^0(\ek)} \;,
\earr \label{eq.fek16}
\ee
where
\be
\barr{l}
\Delta_0(\ek) = \l(q^v - q^{hv}\r) + \dis\f{X\;\xi^2}{1-\xi^2} \l[(\ek)^2 -1\r] \;;\\
X = \dis\f{q^{hv}}{2} \l[ \binom{m}{k}^{-1} \binom{m-k-t}{k} -q^{hv}\r]\;.
\earr \label{eq.fek17}
\ee
With these, we have the important result that $\mu_4(\ek) \sim \mu_4^0(\ek)$ if $\Delta \sim 0$ and $\mu_{40} \sim 3$. 

\subsection{Formulas in the finite $N$ limit}

It is important to mention that in practice, the number of sp states $N$ is finite and therefore it is useful to have finite $N$ formulas for the parameters $q^{h}$, $q^v$ and $q^{hv}$. For sake of completeness, we give the formulas here, which follow from the results given in \cite{KM-strn}. With $V(k)$ represented as an EGOE$(k)$, the formula for $q^v$ is
\be
\barr{l}
q^v =  \;{\dis\binom{N}{m}}^{-1} \; \dis\sum_{\nu=0}^{min(k,m-k)}\,
\dis\f{\Lambda^{\nu}(N,m,k)\, \Lambda^{\nu}(N,m,m-k)\,d(N:\nu)}{
\l[\Lambda^0(N,m,k)\r]^2}\;;\\ \\
\Lambda^{\mu}(N^\pr,m^\pr,r) = \dis\binom{m^\pr-\mu}{r}\,\dis\binom{
N^\pr-m^\pr+r-\mu}{r} \;, \\ \\
d(N:\nu) = {\dis\binom{N}{\nu}}^2 -{\dis\binom{N}{\nu -1}}^2\;.
\earr \label{eq.finite1}
\ee
This equation also gives formula for $q^{h}$ by replacing $k$ by $t$ as $H_0(t)$ is represented by an EGOE($t$). Also, $\Lambda^0(N,m,r)$ gives finite-$N$ formula for the correlation coefficient $\xi$; see Eqs.
(A.7)-(A.9) in \cite{MK}. The formula for $q^{hv}$ is
\be
q^{hv} = \dis\frac{\dis\sum_{\nu=0}^{min(t,m-k)}\,
\Lambda^{\nu}(N,m,k)\, \Lambda^{\nu}(N,m,m-t)\,d(N:\nu)}{
\dis\binom{N}{m}\;\Lambda^0(N,m,t)\,\Lambda^0(N,m,k)}\;.
\label{eq.finite2}
\ee

\section{Discussion of results}

Representing $H_0$ as a mean-field operator i.e. $t = 1$, Eq. \eqref{eq.hh} gives $H = H_0(1) + \lambda \, V(k)$. We consider two examples: (a) $m = 8$ fermions distributed in $N = 20$ sp states and (b) $m = 8$ fermions distributed in $N = 20$ sp states, with $2 \leq k \leq m$. We consider the thermodynamic regime defined by $\xi^2=1/2$, in which wavefunctions look alike i.e. there is no basis dependence (we have two basis defined by $H_0(1)$ and $V(k)$ respectively) \cite{Ko-book}.

\begin{figure}
     \centering
         \includegraphics[width=6in,height=2.5in]{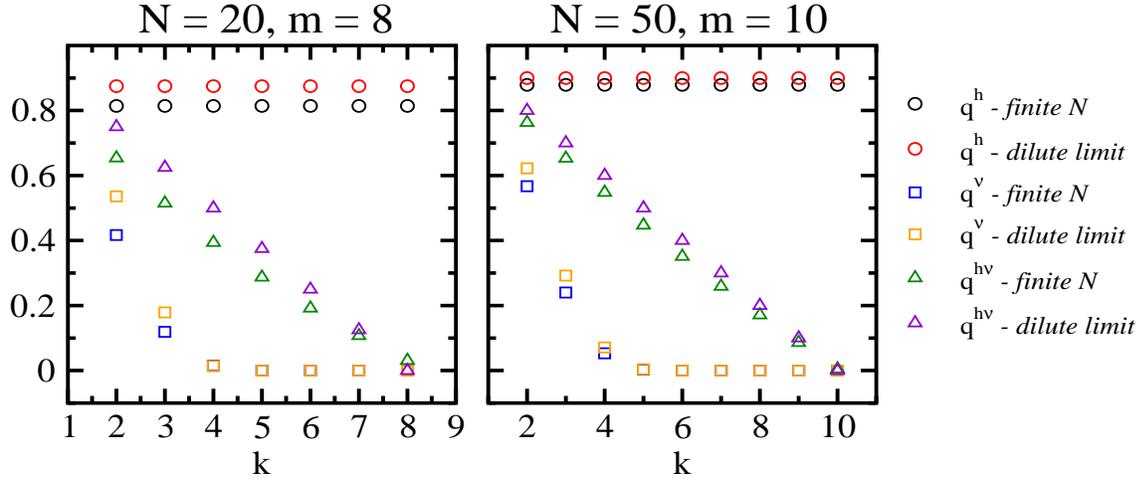}
         \caption{Comparison of parameters $q^h$, $q^v$, $q^{hv}$ in the finite $N$ limit with their respective values in the dilute limit for $N = 20, \, m = 8$ (left panel) and $N = 50, \, m = 10$ (right panel) as a function of rank of interactions $2 \leq k \leq m$. Dilute limit formulas follow from Eqs. (\ref{eq.mpq9}) and (\ref{eq.mpq10}) and finite $N$ formulas follow from Eqs. (\ref{eq.finite1}) and (\ref{eq.finite2}).}
        \label{fig:f1}
\end{figure}

First, we compare the dilute limit and finite $N$ limit results for the $q^h$, $q^v$, $q^{hv}$ parameters in Fig. \ref{fig:f1}. Dilute limit formulas follow from Eqs. (\ref{eq.mpq9}) and (\ref{eq.mpq10}) and finite $N$ formulas follow from Eqs. (\ref{eq.finite1}) and (\ref{eq.finite2}).  Note that $q^h$ is independent of interaction rank $k$. Using $\xi^2=\f{1}{2}$, Eq. (\ref{eq.mpq10}) gives $q^H = (q^h + q^v + 2 q^{hv})/4$.
As can be seen from Fig. \ref{fig:f1}, $q^h$ is independent of $k$, $q^v$ and $q^{hv}$ decrease with increasing $k$ for a given $(N, \,m)$. The finite $N$ results and dilute limit results for $q^h$, $q^v$, $q^{hv}$ are quite close and the difference between the two decreases with increasing $N$, $m$ and $k$. Thus, the dilute limit formulas in Sections 4.1 and 4.2 are quite good. 

Next, we compute the values of $\Delta(\ek)$ defined by Eq. \eqref{eq.fek17} for $\ek = (0, 1, 2)$ for $N = 50, \, m = 10$. These are as follows: $\Delta(\ek) = $ (-0.026, -0.061, -0.157), (-0.079, -0.123, -0.24), (-0.108, -0.143, -0.228), (-0.106, -0.125, -0.166), (-0.090, -0.096, -0.109), (-0.071, -0.069, -0.067), (-0.050, -0.045, -0.036), (-0.027, -0.022, -0.015) and (-0.002, -0.001, -0.001) for $k = 2 - 10$ respectively. These show that the approximation $\mu_4(\ek) \sim \mu_4^0(\ek)$ is quite good [see Eq. (\ref{eq.fek16})] with the difference often $ < 10$\%. These differences  will grow for $\ek > 2$ as, in this situation, we need to take higher order terms in the polynomial expansion given by Eq. (\ref{eq.fek2}). To the extent assumption $\mu_4(\ek) \sim \mu_4^0(\ek)$ is valid, $\mu_4(\ek)$ for $F_\kappa(\hat{E})$ is identical to the formula in Eq. (\ref{eq.qbiv11}) for $\mu_4$ of $f_{CqN}$ with $q=q^{hv}$. 

\begin{figure}
     \centering
         \includegraphics[width=6in,height=6in]{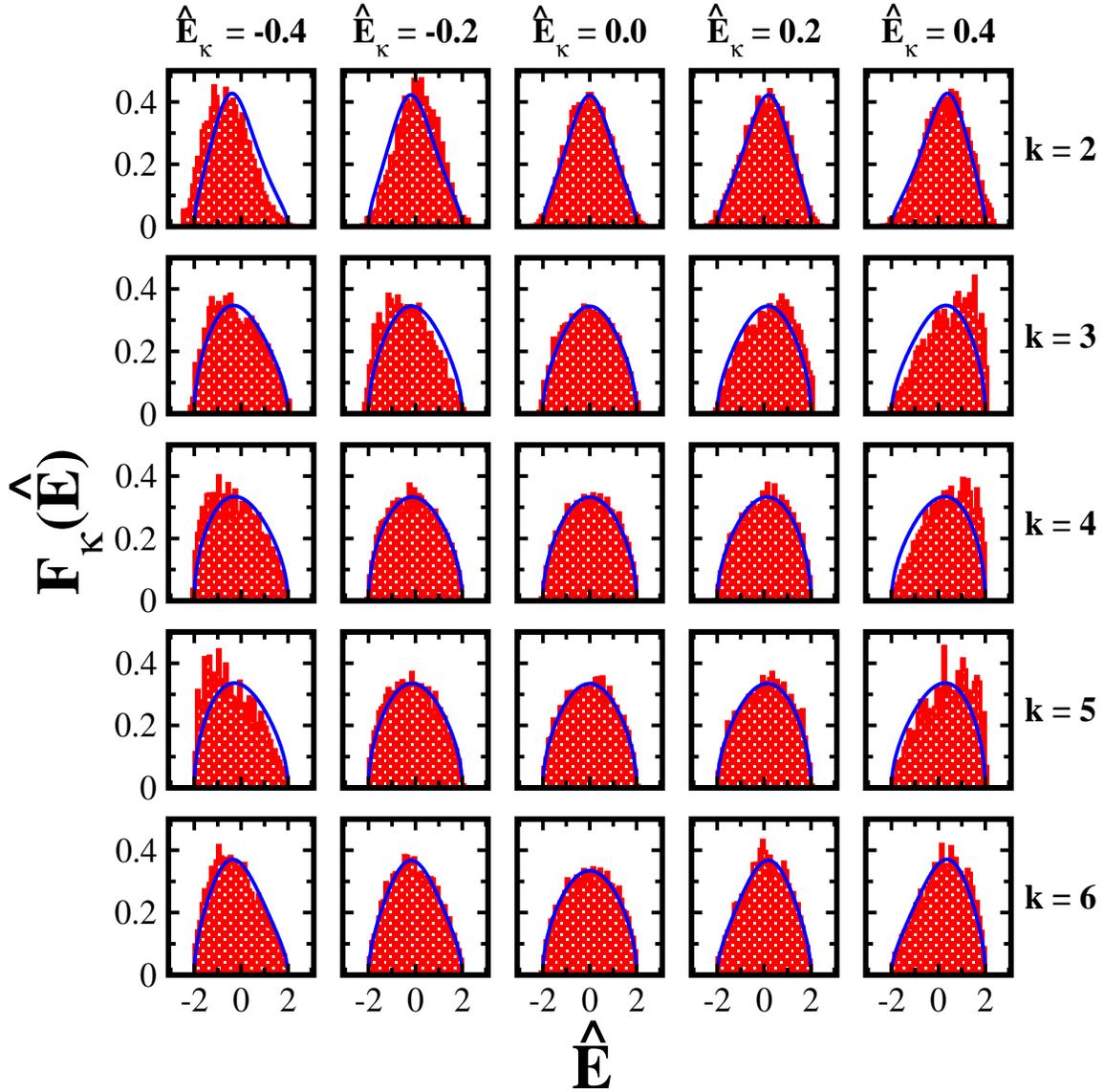}
         \caption{Strength functions $F_\kappa(\hat{E})$ for system of $m = 6$ fermions 
          in $N = 12$ sp levels with $\lambda = 0.5$. We choose fixed $H_0(1)$ and generate
          a 1000 member EGOE$(k)$ ensemble for $V(k)$ defining the system Hamiltonian $H$
          given by Eq. \eqref{eq.hh}. 
          Parameters $k$ and $\ek$ are as indicated in the figure. Note that in the figure, 
          $\hat{E}$ and $\ek$ are normalized eigen and basis state energies (zero centered 
          and scaled by their respective widths). Continuous curves are obtained using conditional 
          $q$-normal densities given by Eq. \eqref{eq.qbiv-3} with parameters $\xi$ and $q$ 
          given by Eqs. \eqref{eq.mpq6} and \eqref{eq.finite2} respectively. See text for further details.}
        \label{fig:fke}
\end{figure}

In the thermodynamic regime ($\xi^2=1/2$), with $t=1$, Eqs. \eqref{eq.mpq9} and \eqref{eq.mpq10} reduce to 
\be
\mu_{40}= 3-\dis\f{1}{m}\;,\;\;\;\mu_{04}=3-\dis\f{(1+k)^2}{4m} + O(1/m^2)\;.
\label{eq.asymp1}
\ee
In deriving the $\mu_{04}$ formula in Eq. (\ref{eq.asymp1}), we have used the following expansion
\be
\binom{m-r}{k}=\dis\f{m^k}{k!}\l[1-\f{1}{m}\l\{kr+\dis\f{k(k-1)}{2}\r\} + O(1/m^2)\r]\;.
\label{eq.asymp2}
\ee
Thus, we have $\mu_{40} \approx \mu_{04} \approx 3$, for $k <<m$; with $(1/m)$ corrections giving close to Gaussian results for $\mu_{40}$ and $\mu_{04}$. However, there will be more deviations with increasing $k$ values. Thus, the equality between $\mu_{PQ}$ and $\mu_{QP}$ will be close in the thermodynamic limit with $k <<m$.
Also, Eq. (\ref{eq.qbiv10}) gives $\gamma_1(\ek)=-(1-q)\ek$ in the thermodynamic region.

\begin{figure}
     \centering
         \includegraphics[width=6in,height=2.25in]{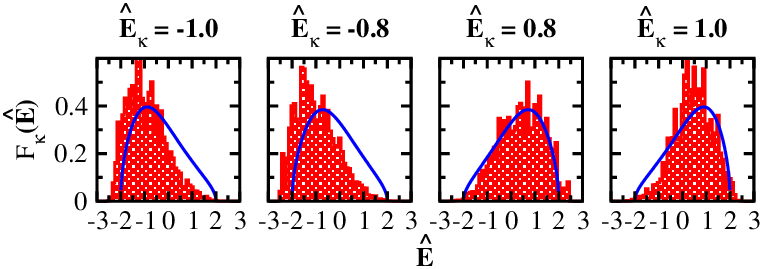}
         \caption{Strength functions $F_\kappa(\hat{E})$ for system of $m = 6$ fermions 
          in $N = 12$ sp levels with $\lambda = 0.5$. We choose fixed $H_0(1)$ and generate
          a 1000 member EGOE$(2)$ ensemble for $V(2)$ defining the system Hamiltonian $H$
          given by Eq. \eqref{eq.hh}.           
          Parameters $\ek$ are as indicated in the figure. Note that in the figure, 
          $\hat{E}$ and $\ek$ are normalized eigen and basis state energies (zero centered 
          and scaled by their respective widths). Continuous curves are obtained using conditional 
          $q$-normal densities given by Eq. \eqref{eq.qbiv-3} with parameters $\xi$ and $q$ 
          given by Eqs. \eqref{eq.mpq6} and \eqref{eq.finite2} respectively.}
        \label{fig3}
\end{figure}

Comparing the moments of strength function $F_\kappa(\hat{E})$ derived in Section 4.2 with the corresponding moments of conditional $q$-normal distribution $f_{CqN}(\hat{E}|\ek ; \xi , q^{hv})$ derived in Section 3 shows the following:

\begin{itemize}

\item Strength functions $F_\kappa(\hat{E})$ show linear variation of the centroids with $\ek$ and the slope is given by the correlation coefficient $\xi$ as seen from Eq. \eqref{eq.fek6}. In addition, Eq. \eqref{eq.fek7} shows there is the constancy of variances i.e., variances are independent of $\ek$. These results are in complete agreement with the properties of the first two moments of $f_{CqN}(\hat{E}|\ek ; \xi , q^{hv})$ given by Eqs. \eqref{eq.qbiv-5} and \eqref{eq.qbiv8} respectively.  

\item Turning to the third reduced moment, as seen from Eqs. (\ref{eq.qbiv10}) and (\ref{eq.fek10}), the $\gamma_1$ is no longer zero for $\ek \neq 0$. It is easy to see that for $\ek$ negative, $\gamma_1$ is positive and therefore $F_\kappa(\hat{E})$ will be skewed in the positive direction. Similarly, for $\ek$ positive, $\gamma_1$ is negative and hence $F_\kappa(\hat{E})$ will be skewed in the negative direction. Formula for the third moment for $F_\kappa(\hat{E})$ [Eq. \eqref{eq.fek10}] is same as the formula for third moment for $f_{CqN}(\hat{E}|\ek ; \xi , q^{hv})$ [Eq. \eqref{eq.qbiv10}]. 

\item For $\ek = 0$, Eq. (\ref{eq.qbiv12}) gives
\be
\gamma_2(\ek)=(1-q)\l[\dis\frac{\xi^2 (1+q)}{1-\xi^2} -1\r]\;.
\label{eq.g20}
\ee
Thus, excess parameter $\gamma_2(\ek) = q (1 - q)$ in the thermodynamic region and therefore, it is always positive. 
Moreover, the fourth reduced moment for $F_\kappa(\hat{E})$ with $q=q^{hv}$ [Eq. \eqref{eq.fek16}] is same as the one from $f_{CqN}(\hat{E}|\ek ; \xi , q^{hv})$ [Eq. \eqref{eq.qbiv12}] for $\ek$ not larger than 2. 

\end{itemize}

Thus, the formulas for the lowest four moments show that strength functions $F_\kappa(\hat{E})$ follow $f_{CqN}(\hat{E}|\ek ; \xi , q^{hv})$ in general.

It is well known that for $\lambda$ small and $t=1$ in in Eq. \eqref{eq.hh} \cite{Ko-book}, strength functions take Breit-Wigner (BW) form [and this extends for any $t$ of $H_0(t)$]. Note that for $\lambda=0$, strength functions are delta functions each located at $\ek$ and change to BW form quickly with increase in the value of $\lambda$. After some value of $\lambda$, the BW form changes to $f_{CqN}$ form. Therefore, for the applicability of $f_{CqN}$ form for the strength functions, clearly $\lambda$ should be sufficiently large. As the thermalization region is defined by $\xi^2=1/2$, this can be used to determine the $\lambda$ value that is sufficiently large for a given $k$. 

For random matrix Hamiltonian $H$, defined in Eq. \eqref{eq.hh} choosing $\lambda = 0.5$, for a system of $m = 6$ fermions distributed in $N = 12$ sp states, we generate a 1000 member EGOE$(k)$ for $V(k)$ operator and choose $H_0$ operator to be defined by fixed sp energies $i+1/i$; $i = 1$, 2, $\ldots, N$. Here, $m$-fermion matrix dimension is $d = 924$. Choosing $\ek = 0.0$, $\pm 0.1$ and $\pm 0.2$, we numerically construct energy distribution of the ensemble averaged strength functions $F_\kappa(\hat{E})$ using Eq. \eqref{eq.fke}. These are shown as histograms (red) in Fig. \ref{fig:fke} as a function of interaction rank $2 \leq k \leq m$. Numerical histograms are compared with theoretical continuous curves (blue) obtained using formula for $f_{CqN}(\hat{E}|\ek ; \xi , q^{hv})$ given in Eq. \eqref{eq.qbiv-3}, with parameters $\xi$ and $q^{hv}$ respectively given by Eqs. \eqref{eq.mpq6} and \eqref{eq.finite2}. Similarly, Fig. \ref{fig3} shows the energy distribution of the ensemble averaged strength functions $F_\kappa(\hat{E})$ for $k = 2$ with $\ek = \pm 0.8$ and $\pm 1.0$. As can be seen from these figures, the theory captures the trends seen in the numerics. However, there are deviations between numerics and theory due to the following reasons: (a) the $H_0$ operator in theory is chosen to be an independent EGOE($1$) while in numerics, we choose $H_0(1)$ to be fixed; and (b) in the example chosen, $(m = N/2)$ and not $(m/N \to 0)$ as needed in the dilute limit. Accounting for these differences requires a larger example (large $N$ and large $m$) for which numerics are prohibitive; the systems shown in Fig. \ref{fig:f1} are not practical as the matrix dimensions are far too large. Also, the deviations increase as increasing and decreasing $\ek$. We can also see that the strength functions are skewed in positive direction for $\ek$ negative and vice-versa. The effect of positive excess parameter is also seen in the plots. Thus, the strength functions $F_\kappa(\hat{E})$ follow $f_{CqN}(\hat{E}|\ek ; \xi , q^{hv})$. 

As $F_\kappa(\hat{E})$ follow $f_{CqN}(\hat{E}|\ek ; \xi , q^{hv})$, it is possible to write NPC and $S^{info}$ in wavefunctions as integrals involving $F_\kappa(\hat{E})$. These chaos markers are defined as follows,
\be
\barr{rcl}
\mbox{NPC}(E) & = & \l[ \dis\frac{1}{d \cdot \rho_2(E)} \dis\sum_{\alpha \in \kappa,\beta \in E^\pr; \, \kappa, E^\pr} \,\l|C^{E^\pr,\beta}_{\kappa,\alpha}\r|^4 \;\delta(E-E^\pr) \r]^{-1} \;,\\
\\
S^{info}(E) & = & - \dis\frac{1}{d \cdot \rho_2(E)}\dis\sum_{\alpha \in \kappa,\beta \in E^\pr; \, \kappa, E^\pr} \,\l|C^{E^\pr,\beta}_{\kappa,\alpha}\r|^2\; \ln \l|C^{E^\pr,\beta}_{\kappa,\alpha}\r|^2 
\;\delta(E-E^\pr) \;.
\earr \label{eq.npc1}
\ee
In Eq. (\ref{eq.npc1}), note that in these formulas $F_\kappa(E)$ with all $E_\kappa$ will enter.  The overlaps $C^{E^\pr,\beta}_{\kappa,\alpha}$ are defined in Eq. \eqref{eq.ovlp}. The integral formula for the ensemble averaged NPC is \cite{KS,Ko-book}
\be
\mbox{NPC}(E) = \dis\f{d}{3} \l[ \dis\int dE_\kappa \;\dis\f{\rho_1(E_\kappa)\;\l\{F_\kappa(E)\r\}^2}{\l\{\rho_2(E)\r\}^2}\right]^{-1}\;.
\label{eq.npc2}
\ee
This is derived as follows (for brevity we will drop the $\alpha$ and $\beta$ labels in Eq. (\ref{eq.npc1})). First write $|C^E_\kappa|^2$ as $|{\cal C}^E_\kappa|^2 \;\overline{|C^E_\kappa|^2}$ where $|{\cal C}^E_\kappa|^2 = |C^E_\kappa|^2/\overline{|C^E_\kappa|^2}$ is the locally renormalized strength and $\overline{|C^E_\kappa|^2}$ is the smooth part of $\overline{|C^E_\kappa|^2}$ (locally/ensemble averaged). Assuming GOE behavior for strength fluctuations (Porter-Thomas law) will give 
$\overline{|{\cal C}^E_\kappa|^4}=3$ (the overline represents ensemble average here). Therefore, $\overline{|C^E_\kappa|^4} = 3 \, \{\overline{|C^E_\kappa|^2} \}^2$. Now writing $\overline{|C^E_\kappa|^2}$ in terms of strength functions and state density using Eq. (5) and replacing the sum over $\kappa$ by integral, i.e. $\sum_\kappa = \int\,d \cdot \rho_1(E_\kappa)\,dE_\kappa$ will give Eq. (\ref{eq.npc2}).

As $E$ and $E_\kappa$ are zero centered, using $\hat{E}=E/\sigma_H$ and $\ek=E_\kappa/\sigma_{H_0}$, we can rewrite Eq. (\ref{eq.npc2}) in terms of $f_{qN}$ and $f_{CqN}$ by replacing $\rho_1(\ek) \to f_{qN}(\ek|q^h)$, $\rho_2(\hat{E}) \to f_{qN}(\hat{E}|q^H)$ and $F_\kappa(\hat{E}) \to f_{CqN}(\hat{E}|\ek;\,\xi , q^{hv})$,
\be
\mbox{NPC}(\hat{E}) = \dis\f{d}{3}\;\l[\dis\int_{-\f{2}{\sqrt{1-q_0}}}^{\f{2}{\sqrt{1-q_0}}} d\ek\;\dis\f{f_{qN}(\ek|q^h)\,\l[f_{CqN}(\hat{E}|\ek;\,\xi , q^{hv})\r]^2}{\l[f_{qN}(\hat{E}|q^H)\r]^2}\r]^{-1}\;.
\label{eq.npc3}
\ee
Here, $q_0$ is the minimum of $(q^h, q^{hv}, q^H)$. A similar integral formula can be written for the $S^{info}$ defined in Eq. \eqref{eq.npc1}. Numerical results obtained using Eq. (\ref{eq.npc3}) for NPC and $S^{info}$ in wavefunctions are reported in \cite{PLA-21} assuming $q$'s in Eq. (56) are all same. Thus, strength functions determine the generic wavefunction structure in many-body quantum systems.

\section{Conclusions and future outlook}

Analytical formulas in Section 4.2 for the lowest four moments of the strength functions, when compared with the formulas from $f_{CqN}$ given in Section 3 show that the strength functions $F_\kappa(\hat{E})$ for quantum many-body systems with $k$-body interactions follow conditional $q$-normal distributions $f_{CqN}(\hat{E}|\ek ; \xi , q^{hv})$ and therefore the remarkable result that strength functions are well represented by conditional $q$-normal distributions. Some numerical results are also presented in Section 5 to justify the approximations needed for the validity of this result. It is important to stress that the strength functions contain all the information about wavefunction structure as seen clearly from Sections 2.2 and 4.2. Also, we have ruled out the possibility of constructing strength functions directly from the bivariate $q$-normal distribution $\rho(E_\kappa,E)$ as discussed in Section 4.1. Numerical results in \cite{PLA-21} suggest that the general structure given by EGOE is equally valid for bosonic ensembles.

Importantly, for the correct description of long time behavior of fidelity decay, a cut-off on both sides of the strength functions is needed \cite{Lea}. Similarly, in the study of nuclear level densities, it is necessary to include cut-off on both sides of Gaussian partial densities \cite{Zel, Chang, zel2, Senkov}. Unlike a semi-circle, Gaussian has no natural cut-off and therefore for $k << m$, it is necessary to introduce a cut-off artificially. In this context, it is important to note that $f_{CqN}(\hat{E}|\ek ; \xi , q^{hv})$ representing $F_\kappa(\hat{E})$ has natural cut-off at $\hat{E} = \pm 2/\sqrt{1-q}$ (note that for real systems, $q \neq 1$ as distributions always have a small value for the excess parameter). Therefore, representing $F_\kappa(\hat{E})$ by conditional $q$-normal distribution $f_{CqN}(\hat{E}|\ek ; \xi , q^{hv})$ will be appropriate for the study of the long time behavior of fidelity decay. This may also give a method to determine the $q^{hv}$ value. Similarly, $f_{qN}(\hat{E}|q)$ form for partial densities will give a natural cutoff to be used in level density studies. These two problems will be investigated further in future.

Going beyond the present analytical results given in Sections 3-4 and numerical investigations presented in Section 5, it is necessary to examine and solve the following problems for a more complete description of strength functions in quantum many-particle systems:

\begin{itemize}

\item For describing $F_\kappa(E)$ for all $\lambda$ values, we need to incorporate the BW or BW-like form in $f_{CqN}$. BW-like form appears for small $\lambda$ (weak coupling limit) \cite{Ko-book,Zelea}. 

\item In the conditional $q$-normal definition, the $q$ value in $f_{CqN}$ and the two marginals are same. However, in practice they will not be same (see discussion about $q^{hv}$, $q^{h}$ and $q^v$ in Section 5). At present, we are not aware of a bivariate $q$-normal with different $q$'s for the two marginal densities and the $h$ function given in Section 2.3. This is an important gap in proper representation of strength functions. 

\item In reality, we have $H = h(1) +\sum_{k=1}^{k_{max}} \lambda_k V(k)$ with $k_{max}=3$ or 4; note that $h(1)$ is the mean-field one-body part. This situation needs to be explored (in \cite{Ko-book} there is some discussion for $k_{max}=3$). 

\item Analytical formulas in Section 4 are valid only for fermions. It is more complex to derive the formulas for bosons. It is likely that the $N \rightarrow -N$ and $N \rightarrow m$ symmetries may apply, as used successfully in the past in many examples to obtain results for bosonic systems \cite{Ko-book} from the formulas derived for the fermionic systems. 

\end{itemize}

\section{Acknowledgments}

Thanks are due to N.D. Chavda for useful discussions and correspondence. Thanks are also due to R. Sahu for help in preparing the manuscript. M. V. acknowledges financial support from UNAM/DGAPA/PAPIIT research grant IA101719 and CONACYT project Fronteras 10872.


\section{References}

\begin{thebibliography}{99}

\bibitem{KS} V.K.B. Kota and R. Sahu, Phys. Rev. E {\bf 64}, 016219 (2001).

\bibitem{Ko-book} V.K.B. Kota, Embedded Random Matrix Ensembles in Quantum Physics (Springer, Heidelberg, 2014).

\bibitem{Bo-17} F. Borgonovi, and F. M. Izrailev, AIP Conference Proceedings {\bf 1912}, 020003 (2017).

\bibitem{Ga-18}  M. A. Garcia-March {\it et. al.}, New J. Phys. {\bf 20}, 113039 (2018).

\bibitem{Vil-20} D. Villase{\~n}or {\it et. al.}, New J. Phys. {\bf 22}, 063036 (2020).

\bibitem{To-17} E. J. Torres-Herrera, L. F. Santos, Phil. Trans. R. Soc. A {\bf 375}, 20160434 (2017). 

\bibitem{Ma-14} E. J. Torres-Herrera, M. Vyas and L. F. Santos, New J. Phys. {\bf 16}, 063010 (2014).

\bibitem{Ri-08} M. Rigol, V. Dunjko and M. Olshanii, Nature {\bf 452}, 854 (2008).

\bibitem{Ri-16} L. D'Alessio, Y. Kafri, A. Polkovnikov and M. Rigol, Advances in Physics {\bf 65}, 239 (2016).

\bibitem{HMKC} S. K. Haldar, N. D. Chavda, Manan Vyas, and V. K. B. Kota, J. Stat. Mech: Theor. Expt. {\bf 2016}, 043101 (2016).

\bibitem{Zelea} F. Borgonovi, F. M. Izrailev, L. F. Santos, and V. G. Zelevinsky, Phys. Rep. {\bf 626}, 1 (2016).

\bibitem{Ta-16} M. Tavora, E. J. Torres-Herrera and L. F. Santos, Phys. Rev. A {\bf 94}, 041603(R) (2016).

\bibitem{PRR-20} M. Niknam, L. F. Santos and D. G. Cory, Phys. Rev. Res. {\bf 2}, 013200 (2020).

\bibitem{Swin-18} B. Swingle, Nat. Phys. {\bf 14}, 988 (2018).

\bibitem{Arul-19} A. Lakshminarayan, Phys. Rev. E {\bf 99}, 012201 (2019).

\bibitem{Sch-19} M. Schiulaz, E. J. Torres-Herrera and L. F. Santos, Phys. Rev. B {\bf 99}, 174313 (2019).

\bibitem{Mal-16} J. Maldacena, S. H. Shenker and D. Stanford, J. High Energ. Phys. {\bf 2016}, 106 (2016).

\bibitem{Lea-PRE-R} F. Borgonovi, F. M. Izrailev and L. F. Santos, Phys. Rev. E {\bf 99}, 010101(R) (2019).

\bibitem{Lea-19} F. Borgonovi, F. M. Izrailev and L. F. Santos, Phys. Rev. E {\bf 99}, 052143 (2019).

\bibitem{fke-e1} V.V. Flambaum, G.F. Gribakin and F.M. Izrailev,  Phys. Rev. E {\bf 53}, 
5729 (1996).

\bibitem{fke-e2} B. Georgeot and D.L. Shepelyansky, Phys. Rev. Lett. {\bf 79}, 4365 (1997).

\bibitem{fke-e3} V.V. Flambaum and F.M. Izrailev,  Phys. Rev. E {\bf 56}, 5144 (1997).

\bibitem{fke-e4} N. Frazier, B.A. Brown and V. Zelevinsky,  Phys. Rev. C {\bf 54}, 1665 (1996).

\bibitem{fke-e5} W. Wang, F.M. Izrailev and G. Casati, Phys. Rev. E {\bf 57}, 323 (1998).

\bibitem{fke-e6} Ph. Jacquod, I. Varga,  Phys. Rev. Lett. {\bf 89}, 134101 (2002).

\bibitem{fke-e7} D. Angom, S. Ghosh and V.K.B. Kota,  Phys. Rev. E {\bf 70}, 016209 (2004).

\bibitem{MF} K. K. Mon and J.B. French, Ann.  Phys. (N.Y.) {\bf 95}, 90 (1975).

\bibitem{Br-81} T. A. Brody, J. Flores, J. B. French, P. A. Mello, A. Pandey, and S. S. M. Wong, Rev. Mod. Phys. {\bf 53}, 385 (1981).

\bibitem{Ki-1} A. Kitaev, A simple model of quantum holography (part 1), talk at KITP, http://online.kitp.ucsb.edu/online/entangled15/ kitaev/.

\bibitem{Ki-2} A. Kitaev, A simple model of quantum holography (part 2), talk at KITP, http://online.kitp.ucsb.edu/online/entangled15/kitaev2/.

\bibitem{Sa-93} S. Sachdev and J. Ye, Gapless Spin-Fluid Ground State in a Random Quantum Heisenberg Magnet, Phys. Rev. Lett. 70, 3339 (1993).

\bibitem{Gu-20} Y. Gu, A. Kitaev, S. Sachdeva and G. Tarnopolsky, J. High Energ. Phys. {\bf 2020}, 157 (2020).

\bibitem{Ver-a} Y. Jia and J. J. M. Verbaarschot, J. High Energ. Phys. {\bf 2020}, 1 (2020).

\bibitem{Ver-b} A. M. Garcia-Garcia, T. Nosaka, D. Rosa and J. J. M. Verbaarschot, Phys. Rev. D {\bf 100}, 026002 (2019).

\bibitem{Verba-1} A. M. Garcia-Garcia and J. J. M. Verbaarschot, Phys. Rev. D {\bf 96}, 066012 (2017).

\bibitem{Ver-d} A. M. Garcia-Garcia and J. J. M. Verbaarschot, Phys. Rev. D {\bf 94}, 126010 (2016).

\bibitem{MK} Manan Vyas and V.K.B. Kota, J. Stat. Mech. {\bf 2019}, 103103 (2019).

\bibitem{Ismail} M. E. H. Ismail, D. Stanton, and G. Viennot, Europ. J. Combinatorics {\bf 8}, 379 (1987).

\bibitem{Verba-2} Y. Jia and J. J. M. Verbaarschot, JHEP {\bf 7}, 193 (2020).

\bibitem{MK-new} M. Vyas and V.K.B. Kota, J. Stat. Mech. {\bf 2020}, 093101 (2020).

\bibitem{Sza-1} P. J. Szabowski, Electronic Journal of Probability {\bf  15}, 1296 (2010).

\bibitem{PLA-21} P. Rao and N. D. Chavda, Phys. Lett. A {\bf 399}, 127302 (2021).

\bibitem{CK} N.D. Chavda and V.K.B. Kota, Ann. Phys. (Berlin) {\bf 529}, 1600287 (2017).

\bibitem{Ko-01} V.K.B. Kota, Phys. Rep. {\bf 347}, 223 (2001).

\bibitem{Fr-71} J. B. French and S. S. M. Wong, Phys. Lett. B {\bf 33}, 449 (1970).

\bibitem{Bo-71} O. Bohigas and J. Flores, {\it ibid.} {\bf 34}, 261 (1971); {\bf 35}, 383 (1971).

\bibitem{BW} L. Benet, T. Rupp, and H. A. Weidenm\"{u}ller, Ann. Phys. (N.Y.)  {\bf
292}, 67 (2001).

\bibitem{BW-rev} L. Benet and H. A. Weidenm\"{u}ller, J. Phys. A: Math. Gen. {\bf 36}, 3569 (2003).

\bibitem{Small} R.A. Small and S. Mueller, Ann. Phys. (NY), {\bf 356}, 269 (2015).

\bibitem{Alhassid-review} Y. Alhassid, Rev. Mod. Phys. {\bf 72}, 895 (2000).

\bibitem{Pap-review} T. Papenbrock and H. A. Weidenm\"{u}ller, Rev. Mod. Phys. {\bf 79}, 997 (2007).

\bibitem{Gu-review} T. Guhr, A. M\"{u}ller-Groeling, and H. A. Weidenm\"{u}ller, Phys. Rep. {\bf 299}, 189 (1998).

\bibitem{Szb-2} P. J. Szabowski, Statistics \& Probability Letters {\bf 106}, 65 (2015).

\bibitem{steve} S. Tomsovic, PhD Thesis University of Rochester, Rochester, New York (1986).

\bibitem{DFW} J.P. Draayer, J.B. French and S.S.M. Wong, Ann. Phys. (N.Y.) {\bf 106}, 472 (1977).

\bibitem{KH-10} V.K.B. Kota and R.U. Haq, Spectral Distributions in Nuclei and Statistical Spectroscopy (World Scientific, Singapore, 2010).

\bibitem{KM-strn} V.K.B. Kota and Manan Vyas, Ann. Phys. (N.Y.) {\bf 359}, 252 (2015).

\bibitem{Lea} M. Tavora, E. J. Torres-Herrera, L. F. Santos, Phys. Rev. A {\bf 95}, 013604 (2017).

\bibitem{Zel} R. Sen'kov and V. G. Zelevinsky, Phys. Rev. C {\bf 93}, 064304 (2016).

\bibitem{Chang} F.S. Chang, J.B. French and T.H. Thio, Ann. Phys. (NY) {\bf 66}, 137 (1971).

\bibitem{zel2} M. Horoi, J. Kaiser and V. Zelevinsky, Phys. Rev. C {\bf 67}, 054309 (2003).

\bibitem{Senkov} R. A. Senkov and M. Horoi, Phys. Rev. C {\bf 82},  024304 (2010).

\end{thebibliography}
\end{document}